\documentclass[prl,aps,onecolumn,amsmath,amssymb,floatfix,
superscriptaddress]{revtex4}

\usepackage[dvips]{graphics}

\usepackage{xcolor}

\definecolor{dred}{rgb}{0,0,0.6}

\begin{document}

\title{Relativistic Anyon Beam: Construction and Properties}

\author{Joydeep Majhi}

\affiliation{Physics and Applied Mathematics Unit, Indian Statistical
Institute, 203 Barrackpore Trunk Road, Kolkata-700 108, India}

\author{Subir Ghosh}

\affiliation{Physics and Applied Mathematics Unit, Indian Statistical
Institute, 203 Barrackpore Trunk Road, Kolkata-700 108, India}

\author{Santanu K. Maiti}

\affiliation{Physics and Applied Mathematics Unit, Indian Statistical
Institute, 203 Barrackpore Trunk Road, Kolkata-700 108, India}

\begin{abstract}

Motivated by recent interest in photon and electron vortex beams, we propose 
the construction of a relativistic anyon beam. Following Jackiw and Nair 
[Phys. Rev. D~\textbf{43}, 1933 (1991)] we derive explicit form of 
relativistic plane wave solution of a single anyon. Subsequently we construct 
the planar anyon beam by superposing these solutions. Explicit expressions 
for the conserved anyon current are derived. Finally, we provide expressions 
for the anyon beam current using the superposed waves and discuss its 
properties. We also comment on the possibility of laboratory construction 
of anyon beam.

\end{abstract}

\maketitle

We propose construction of a relativistic beam of anyons in a plane. 
Anyons are planar excitations with arbitrary spin and statistics. The procedure 
is similar to 3D optical (spin $1$) and electron (spin $1/2$) vortex beam.

Vortex beams, of recent interest, carrying intrinsic orbital angular momentum are 
non-diffracting wave packets in motion. In 3D, vortex beams for photons and electrons 
were proposed by Durnin {\em et al.}~\cite{dur} and by Bliokh {\em et al.}~\cite{bl1} 
respectively (see~\cite{bl3}). Properties and experimental signatures of twisted optical 
and electron vortex beam were studied respectively in~\cite{allen} (for spinless electron \cite{bl1}) and 
\cite{silenko} (spinor electron \cite{bl1}). Wave packets were constructed out of free 
plane wave, orbital angular momentum solutions having vorticity. In 3D, non-trivial Lie 
algebra of rotation generators along with finite dimensional representations of angular momentum eigenstates 
restricts the angular momentum to quantize in units of $\hbar/2$. Absence of such an algebra in 2D allows 
excitations (anyons~\cite{wil1,lein}) with arbitrary spin and statistics with
infinite dimensional representations. Here, we construct directed transversely localized anyon 
wave packet or wave beam in 2D. 

Geometry imposes a qualitative difference between monoenergetic wave packet 
and wave beam both in 3D and 2D. In 3D the wave packet is described by three 
discrete quantum numbers as it is localized in three directions whereas 
ideally an energy dispersionless wave packet where wave components having 
identical momentum along the direction of propagation will be completely 
delocalized along that direction and will constitute a beam~\cite{dur,bl1}. 
But, in 2D a monoenergetic wave beam will necessary have dispersion in momentum
along propagation direction and hence cannot be completely delocalized. The 
delocalization along propagation direction will be more for narrower 
angle of superposition.

Apart from earlier interest~\cite{wil1}, graphene  also involves anyons~\cite{graph}. Non-Abelian 
anyons~\cite{nonab gr} are touted as theoretical building blocks for topological fault-tolerant
quantum computers~\cite{kit}. An exact chiral spin liquid with 
non-Abelian anyons has been reported~\cite{spin}. The direct 
observational status of anyons has shown promising development~\cite{expt}. 

The {\em minimal} field theoretic and relativistic model of a single 
anyon was constructed by Jackiw and Nair (JN)~\cite{jn}. JN anyon is the most suitable 
one that serves our purpose since, being first order in spacetime derivatives, it closely resembles 
the Dirac equation used by~\cite{bl1}. A major difference is that unitary representation for  
arbitrary spin JN anyon requires an {\em infinite} component wave function~\cite{bin},  
(to maintain covariance), along with subsidiary constraints that 
restrict the number of independent components to a single one~\cite{jn}. 

We propose: (i) an explicit form of the single anyon solution of the JN anyon 
equation~\cite{jn} (which is a new result), (ii) anyon conserved current (again a new result),
(iii) anyon wave packet in 2D, (iv) anyon wave packet in the conserved current that constitutes 
the anyon beam, (iv) anyon charge and current densities for 
the anyon beam numerically since closed analytic expressions could not 
be obtained, (v) possible laboratory setup for observing anyon beams, and (vi) finally,
future prospects. The new results (i, ii) lead us to our principal outputs 
(iii, iv) and hopefully (v).

\noindent
{(i) \bf{Jackiw-Nair anyon equation and its solution:}} We start with 
a familiar system, free spin one particle in $2+1$-dimensions~\cite{jn}. 
The dynamical equation in co-ordinate and momentum space ($i\partial_a = p_a$) 
is given by
\begin{equation}
\partial _a \epsilon^{abc}F_c \pm mF^b=0;~~
(p \cdot j)^a_b F^b+msF^a = 0.
\end{equation}
The solution of the three-vector $F^a,~a=0,1,2$ (we use Minkowski 
metric $\eta_{\mu\nu}=diag(1,-1,-1)$) is given by
\begin{equation}
F^a(p)=\frac{1}{\sqrt{2}}\Bigg[\begin{bmatrix}
0 \\
1 \\
i \\ 
\end{bmatrix}
+\frac{p^x + i p^y}{m(E+m)}\begin{bmatrix}
E+m \\
p^x \\
p^y \\ 
\end{bmatrix}
\Bigg] \sqrt{\frac{m}{E} }.
\label{ff}
\end{equation}
The same $F^a$ can also be expressed as a Lorentz boosted form
\begin{equation}
F^a(p)= B^a_b(p) N^b_c F^c_0(p),~~
N^b_c = \frac{1}{\sqrt{2}}\left[ 
\begin{tabular}{ccc}
$\sqrt{2}$ & 0 & 0 \\
0 & 1 & $i$ \\
0 & -$i$ & $i$ 
\end{tabular}
\right]
\label{f1} 
\end{equation}
where the boost is expressed by the generators $j^a$ 
\begin{eqnarray}
B(p)  =  e^{i\Omega_a(p)j^a}; 
[B(p)]^a_b=[B^{-1}(p)]_b^a \nonumber \\= \delta ^a_b - \frac{(p^a + \eta ^a m)(p_b+\eta_b m)}{m(p \cdot \eta +m)}+ \frac{2p^a\eta_b}{m} 
\end{eqnarray}
with $\eta_a=(1,0,0)$. It is straightforward to check that $F^a$ 
describes a spin one particle ($s=1$) of mass $m$.

This construction has been extended in an elegant way to the JN anyon 
equation~\cite{jn} to describe an anyon of arbitrary spin  $s=1-\lambda$, 
whose dynamics in  momentum space is given by,
\begin{equation}
P\cdot(K+j)_{an \hspace{0.1cm}a'n'} f^{a'}_{n'} + ms f_{an}=0,~~(D_af^a)_n =0,
\label{an1}
\end{equation}
($D^a_{nn'}=\epsilon^a_{~bc}P^bK^c_{nn'}$) where the second 
equation is the subsidiary (constraint) relation. For $\lambda =0$ the 
anyon reduces to the spin one model discussed earlier~\cite{jn}.
The actions of $j^a,K^a$ are given by
\begin{eqnarray}
P\cdot K_{an~a'n'}=P\cdot K_{nn'}\delta_{aa'},~P \cdot j_{an~a'n'}=P\cdot j_{aa'}\delta_{nn'}; 
\nonumber \\ K^a_{nn'}=<\lambda, 
n\mid K^a\mid \lambda, n'>,~(j^a)_{a'a''}=i\epsilon_{a'~a''}^{~a} \nonumber
\end{eqnarray}

Generalizing the spin-$1$ case (\ref{f1}), the free anyon 
solution is formally given by~\cite{jn}
\begin{equation}
f_n^{a(\pm)}(p) = B_{n0}(p) B^a_b (p)N^b_c f^{c(\pm)}(p),
\label{bnn}
\end{equation}
where $B_{nn'}(p),~ B^a_b (p)$ are the spin $\lambda$ and spin $1$ 
representations of the boost transformation respectively and 
$N^b_c$ is the same numerical matrix as in (\ref{f1}).

Exploiting coherent state formalism for $SU(1,1)$ \cite{per,vpn} along 
with the $SU(1,1)\sim SO(2,1)$ connection, we have constructed explicit 
form of $B_{n0}(p)$ (in a matrix representation of $K^a$~\cite{jn} (see 
e.g.~\cite{wy}). Computational details are in {\bf{Suppl. material A}}. 
The free anyon solution is,
\begin{eqnarray}
f^{a+}_n= \left(\frac{2m}{E+m} \right)^\lambda  
\sqrt{\frac{\Gamma(2\lambda+n)}{n!\Gamma(2\lambda)}} \left(\frac{p^x + i p^y}{E+m}\right)^n 
\nonumber \\ \times  \, F^a(p) \,e^{-ip \cdot x}
\label{fan}
\end{eqnarray}
where $F^a(p)$ is same as the spin-$1$ case defined in (\ref{ff}). 
This is our primary result that we exploit to construct the wave packet 
and subsequent anyon beam.

\noindent
{(ii) \bf{Conserved current for single anyon:}} Next we derive the 
conserved probability current for anyon $\partial^\mu j^{(s=1-\lambda)}_\mu =0$ 
where $j_0^s$ is the probability density. Since the anyon model of~\cite{jn} 
is an extension of the spin-$1$ case we can take a cue from the latter where 
the conservation law $\partial^\mu j^{(s=1)}_\mu =0$ reads
{\small \begin{eqnarray}
	& \partial^\mu j^{(s=1)}_\mu =
\partial^0\left[F^{0\dagger}F^0+F^{x\dagger}F^x+F^{y\dagger}F^y\right]
\nonumber \\
	& -\partial^x\left[F^{0\dagger}F^x+F^{x\dagger}F^0\right]  
-\partial^y\left[F^{0\dagger}F^y+F^{y\dagger}F^0\right]=0
\label{cont1}
\end{eqnarray}}
Considering the Fourier transform of the anyon equation of motion 
(\ref{an1}) in position space, a long calculation yields the conserved free (single) 
anyon current $j^{(1-\lambda)}_\mu$,
{\scriptsize
\begin{eqnarray}
&&\partial^0\sum\limits_{n=0}^{\infty} \left[(f_n^{0\dagger}f^0_n + f_n^{x\dagger}f^x_n + 
f_n^{y\dagger}f^y_n) - i(f_n^{y\dagger}K^0_{nn'} f^x_{n'} - f_n^{x\dagger}K^0_{nn'}f^y_{n'})\right]
\nonumber \\ 
&&- \partial^x\sum\limits_{n=0}^{\infty} \left[(f_n^{0\dagger}f^x_n+f_n^{x\dagger}f^0_n) - 
i(f_n^{y\dagger}K^x_{nn'} f^x_{n'} - f_n^{x\dagger}K^x_{nn'}f^y_{n'})\right] \nonumber \\
&&-\partial^y\sum\limits_{n=0}^{\infty} \left[(f_n^{0\dagger}f^y_n+f_n^{y\dagger}f^0_n) - 
i(f_n^{y\dagger}K^y_{nn'}f^x_{n'} - f_n^{x\dagger}K^y_{nn'}f^y_{n'})\right]=0 \nonumber \\
\label{ac}
\end{eqnarray}
}
where we have explicitly shown the summation over $n$, the anyonic index. 
For $\lambda =0$ the current $j_\mu ^{(1-\lambda)}$ reduces to the spin $1$ 
current $j_\mu ^{(1)}$ of (\ref{cont1}). A nontrivial check of the consistency 
of the expressions for anyon current (\ref{ac}) is to substitute 
$f^a_n$ from (\ref{fan}) to yield
\begin{eqnarray}
j^0=(1-\lambda)E/m;~~j^x=(1-\lambda)p^x/m; \nonumber \\
j^y=(1-\lambda)p^y/m ~\rightarrow j^\mu=(1-\lambda)\frac{p^\mu}{m}=s\frac{p^\mu}{m}.
\label{acc}
\end{eqnarray}
For computational details see {\bf Suppl. material B}.

\begin{figure}[ht]
{\centering \resizebox*{7cm}{5.5cm}{\includegraphics{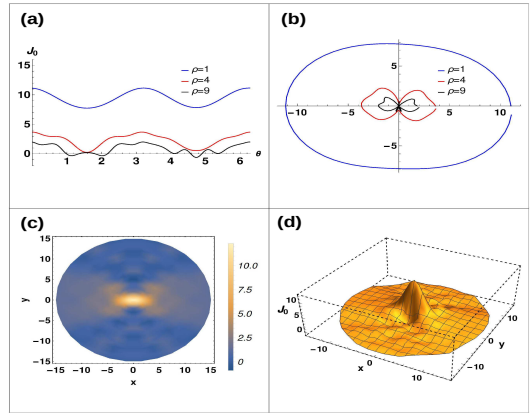}}\par}
\caption{(a) $J^0$-$\theta$ for different
values of $\rho$. (b) Polar plot for $J^0$ with $\theta$ for fixed $\rho$
where the radial distance is the magnitude of $J^0$. (c) Density
plot of $J^0$. (d) $3$D plot of $J^0$. We set $\lambda=0.6$ and the
integration range from $-\pi/2$ to $+\pi/2$.}
\label{fig2}
\end{figure}

\noindent
{(iii) \bf{Anyon wave packet:}} Our aim is to construct the anyon current, 
not for a single anyon as done above, but for an anyon wave packet which 
can be amenable to experimental verification. Let us now construct the 
anyon wave packet that we want to move towards, say, $+x$-direction. 
Since we have superposed plane waves, later figures will reveal that the 
current density has a sharply peaked profile with the $y$-component of 
current density having a comparatively reduced value. Note an important 
difference in geometry between our construction and that of the 
three-dimensional vortex  beam~\cite{bl1}. In the latter case the free 
monoenergetic 
plane wave solutions (to be superposed) are distributed over the surface 
of a right circular cone with identical momentum amplitude in the propagation 
direction. However, for our anyon wave packet, in a planar geometry the 
above is not possible. Instead we use the superposition scheme where the 
azimuthal angle $\phi$ of momenta of the plane wave is integrated 
symmetrically from $\phi_0=-\pi/2$ to $\phi_0=+\pi/2$.
In Fig.~\ref{fig2} we have shown profiles for charge density of anyon 
beam, $J_0^\lambda$, for $\lambda =0.6 \rightarrow s=1-\lambda =0.4$ for 
$\phi_0=\pm \pi/2$.

Hence, considering superpositions of anyon plane waves with fixed 
spin $s=1-\lambda$, fixed energy $E$ and fixed kinetic energy 
$p={\sqrt{(p^x)^2+(p^y)^2}}$, for the special case of integration 
limits $\phi_0=\pm \pi/2$, the superposition $F^a_n$ appears as
\begin{eqnarray}
F_n^a(p,x)={\mathcal {A}}_n\int\limits_{-\pi/2}^{\pi/2}\begin{bmatrix}
\frac{p}{m} e^{i n \alpha}e^{i \alpha} \\
\left(1+\frac{p^2}{M} \cos\alpha \, e^{i\alpha}\right) e^{in\alpha} \\
\left(i+\frac{p^2}{M} \sin\alpha \, e^{i\alpha}\right) e^{in\alpha}  \\ 
\end{bmatrix} \nonumber \\
\left(e^{i(x \cos \alpha + y \sin \alpha )} + e^{i(x \cos \alpha - 
y \sin \alpha )}\right)\, d\alpha
\label{ffn}
\end{eqnarray}
where ${\mathcal {A}}_n = \frac{1}{2} \left(\frac{2m}{E+m}\right)^{\lambda}
\sqrt{\frac{\Gamma(2\lambda+n)}{n!\Gamma(2\lambda)}}\sqrt{\frac{m}{E} }\left(\frac{p}{E+m}\right)^n$, $x \equiv px$, $y \equiv py$ and $M=m(E+m)$. 
The expression is symmetric separately under $x\rightarrow -x$ and $y\rightarrow -y$.

\noindent
{\bf{Conserved  current for anyon wave packet:}} The final analytical 
task is to substitute the anyon wave packet (\ref{ffn}) in 
the expression of the anyon current (\ref{ac}). Since the current components 
are quadratic in the packet wavefunctions $F^a_n$, the final expressions are quite 
long and involved. We have shown only the expression for probability 
density $J^0$ and have relegated $J^x$ and $J^y$ to {\bf{Suppl. material C}} 
together 
with a few computational steps. The cherished form of anyon beam probability 
density, in polar coordinates $x=\rho \cos\theta$, $y=\rho \sin\theta$, is
{\small
\begin{widetext}
\begin{eqnarray}
J^0(\rho, \theta) & = &  \left(\frac{2m}{E+m}\right)^{2\lambda} 
\frac{m}{E} \int\limits_{-\pi/2}^{\pi/2} d\alpha 
\int\limits_{-\pi/2}^{\pi/2} d\beta \left[e^{-2i \rho p \sin(\theta -(\alpha+ \beta)/2) \sin((\alpha-\beta)/2)} 
+ e^{-2i \rho p (\sin(\theta -(\alpha- \beta)/2) \sin((\alpha+\beta)/2)} \right.\nonumber \\ 
& & \left. + e^{-2i \rho p (\sin(\theta +(\alpha- \beta)/2) \sin(-(\alpha+\beta)/2)} +
e^{-2i \rho p (\sin(\theta +(\alpha- \beta)/2) \sin((-\alpha+\beta)/2)}\right] 
\left[\left[1 - e^{-i (\alpha-\beta)} \sigma^2\right]^{-2 \lambda} \right. \nonumber \\ 
& & \left. \left\{\left( \frac{p}{2m} \right)^2 e^{-i( \alpha-\beta)} 
+ \frac{1}{4} \left\{(1+\lambda)(2+ \frac{2p^2}{M}) + \frac{p^4}{M^2} e^{-i (\alpha-\beta)} (\cos(\alpha-\beta) 
+ i \lambda \sin(\alpha-\beta)) \right\} \right\} \right. \nonumber \\ 
& &\left. -\frac{\lambda}{2} \left\{ \left[1 - e^{-i (\alpha-\beta)} \sigma^2\right ]^{-2 \lambda -1 } 
\left(2 + 2\frac{p^2}{M} + i \frac{p^4}{M^2} \sin(\alpha-\beta)e^{-i(\alpha-\beta)}\right) \right\}\right]
\label{J0}
\end{eqnarray}
\end{widetext}
}
where, $\sigma=\left(\frac{p}{E+m} \right)$.

\begin{figure}[ht]
{\centering \resizebox*{7cm}{5.5cm}{\includegraphics{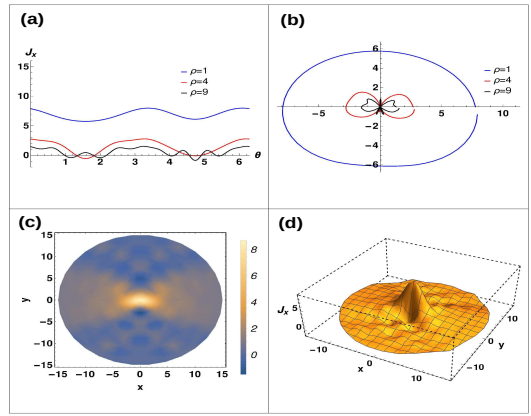}}\par}
\caption{$J^x$ ($\lambda=0.6$) with integration range $\pm \pi/2$, where 
(a)-(d) correspond to the similar meaning as described in Fig.~\ref{fig2}.}
\label{fig5}
\end{figure}

\begin{figure}[ht]
{\centering \resizebox*{7cm}{5.5cm}{\includegraphics{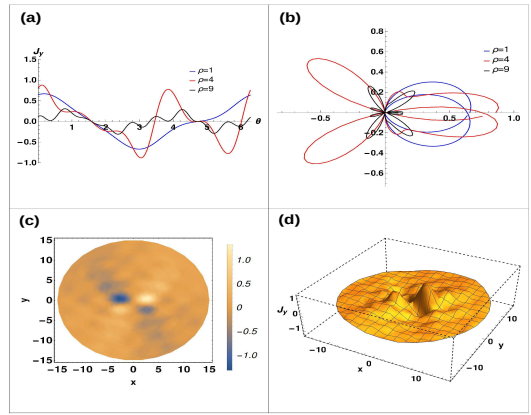}}\par}
\caption{$J^y$ ($\lambda=0.6$) with integration range $\pm \pi/2$, where
(a)-(d) represent the similar meaning as described in Fig.~\ref{fig2}.}
\label{fig6}
\end{figure}

\noindent
{(iv) \bf{Visualizing the anyon beam:}} Unfortunately, closed form 
expressions for the anyon beam current components $J^0$, $J^x$, and $J^x$ 
are not possible to obtain. 
For different choices of $\lambda$ and
$\phi_0$ the profiles of $J^0$ are given in {\bf Suppl. material D}.  
Subsequently, in  Fig.~\ref{fig5} and Fig.~\ref{fig6}, for the above values of 
$s,\phi_0$, we have plotted the profiles of $J^a, ~~a\equiv x,y$ respectively, in three ways: 
a two-dimensional plot of $J^a$  against the polar angle $\theta$ for a few 
values of the (planar) radial distance $\rho$, panel (a) in each group of 
figures. Another two-dimensional graph of the same data as panel (a) with 
magnitude of $J^a$ against $\theta$ for the same values of $\rho$ is shown 
in panel (b). A density plot in co-ordinate plane $x$-$y$ is given in 
panel (c). Finally, a  three-dimensional plot of $J^a$ in co-ordinate plane 
$x$-$y$ is provided in panel (d). Note that in panel (a), each continuous 
curve represents a fixed polar distance in coordinate plane $x$-$y$ with 
the height being a measure of $J^a$ whereas in panel (b), the radial 
distance  is a measure of the intensity of $J^a$. Hence the curves that
are further away from the centre in panel (b) represent points that 
are closer to the coordinate plane $x$-$y$. 

As expected, all the wave packet profiles are symmetric about the abscissa 
since the packets are superposition of plane waves, that are symmetrically
placed about the $x$-axis. Contrasting with the three-dimensional wave
packets~\cite{bl1} it is clear that the planar anyon beams do not
possess a vortex nature since the axial symmetry is manifestly broken 
while constructing a propagating anyon beam. This is also corroborated 
in the figures that do not have any destructive interference at the 
origin, a characteristic feature of vortex beams~\cite{bl1}. Hence 
the anyon beams are characterized by the spin value $s$ of the wave packet, 
which is same as that of individual plane wave single anyon component.

An important observation is that in the cases we have considered, 
$J^0$ is always positive, which has to be the case since it is the 
probability density. But $J^x$ is also positive throughout whereas 
$J^y$ has positive and negative values in equal amount. Furthermore, 
maximum value of $J^y$ is far lower than each of $J^0$ and $J^x$. 
These reflect the nature of our construction of the anyon beam where 
all the plane waves have {\em positive} velocity along $x$-direction 
but have pairwise {\em opposite} (both $+$ve and $-$ve ) velocities 
along $y$-direction. Hence, the anyon beam will predominantly move in 
the positive $x$-direction with the $y$-component effectively canceled out.

\noindent
{(v) \bf{Experimental possibilities:}} Anyons were detected in quantum antidot
experiments~\cite{anyonobs} and in Laughlin quasiparticle interferometer~\cite{obsan}. 
In relation to simulation of high $T_c$ superconductivity by charged anyon fluid~\cite{hos} their 
Josephson frequency has been observed~\cite{ch an} in 2D electrons in high magnetic field.

External electromagnetic field affects    anyon with charge $e$ and magnetic moment (see e.g. \cite{bl1})
\begin{equation}
 M=e\int dA \,(\epsilon_{ik}r^i j_s^k)/\int dA\,j^0_s
\label{mm}
\end{equation}
for  the single anyon current $j_a^s$  
(\ref{cont1}) using (\ref{fan}). We  replace $j^a$ by $J^a$ using the wave packet 
(\ref{ffn}). The nonstationary quantum superposition state (wave packet) can be 
created by exciting matter coherently with an ultrafast laser pulse, which is composed 
of eigenstates spanned by the frequency bandwidth of the laser~\cite{wpacket}.

Possibility of fault tolerant quantum computation by (non-abelian) anyons 
has generated  interest in controlled production of anyonic~\cite{expt}: 
collective anyon excitations from electrons in the Fractional Quantum Hall (FQH)
systems  or from atoms in 1D optical lattices, creation of FQH  effect for 
photons (using 1D or 2D
\begin{figure}[ht]
{\centering
\resizebox*{2cm}{3cm}{\includegraphics{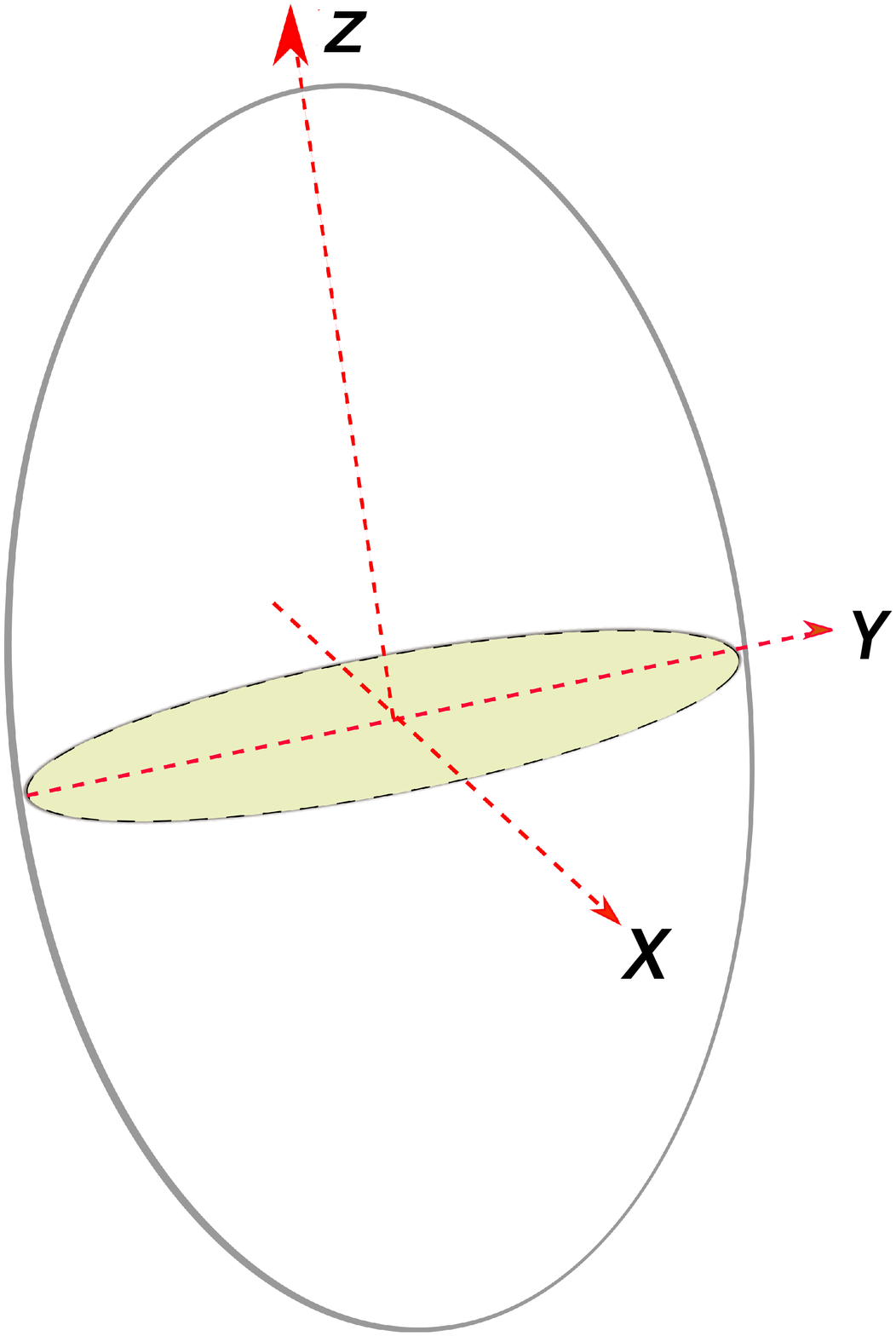}}
$~~~~$\resizebox*{3cm}{3cm}{\includegraphics{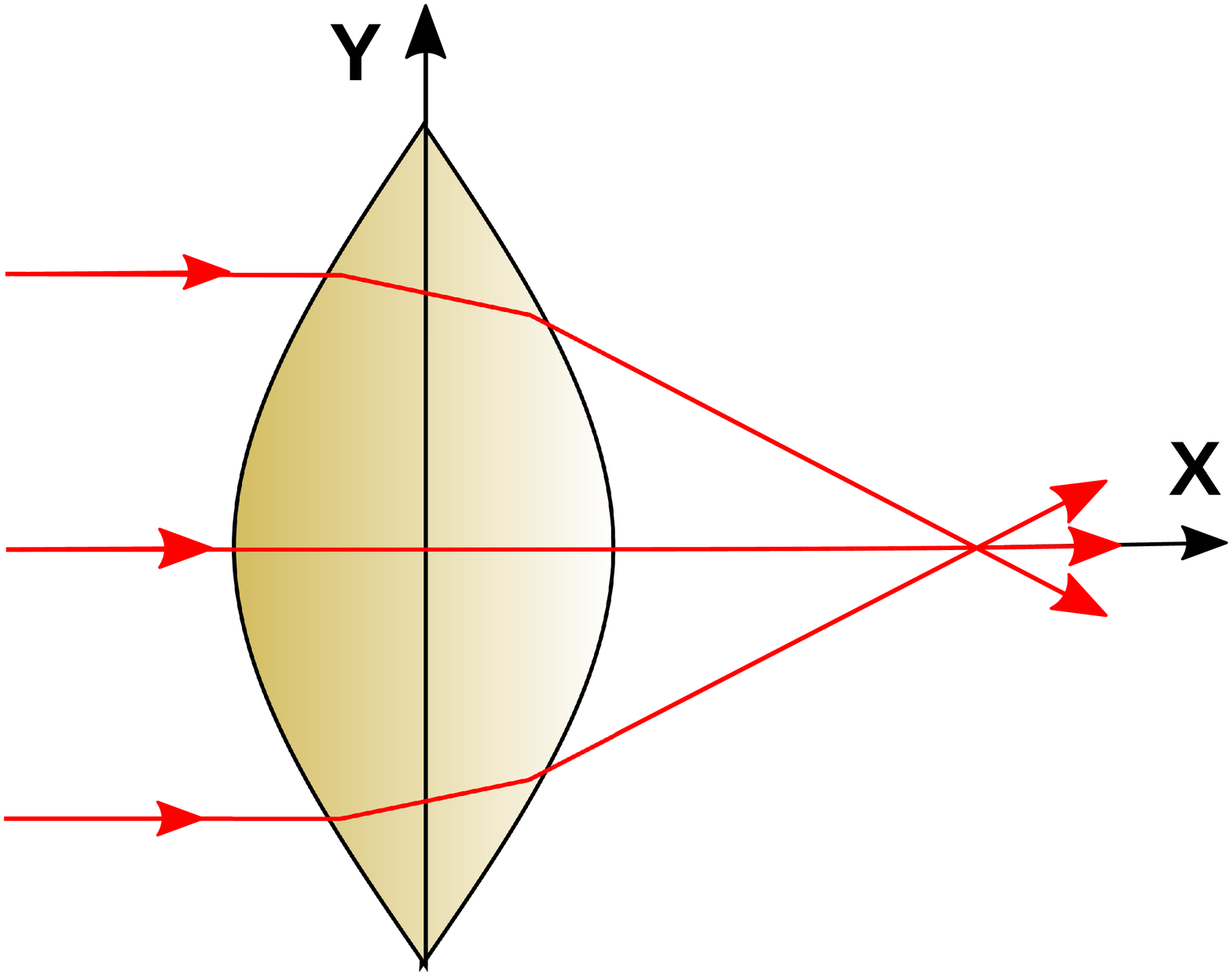}}\par}
\caption{Experimental setup: A slice (projection) of a 3D
convex lens in the $x$-$y$ plane (left diagram) is taken out and shown
separately as thick lens in the right diagram. The red lines show the
superposing wave vectors in $x$-$y$ plane.}
\label{fig7}
\end{figure}
cavity array), using a nonlinear resonator 
lattice subject to dynamic modulation for creating anyons from photons, 
among others.

Topologically ordered many-body states (with quenched kinetic energy) are 
generated with strongly interacting particles in magnetic field. The system 
minimizes  interaction energy forming intricate patterns of long-range 
entanglement as observed in FQH (semiconductor heterojunction, graphene~\cite{12,13} and
van der Waals bilayers~\cite{14}). Formation of Laughlin states in synthetic 
quantum systems (ultracold atoms~\cite{7,8}, photons~\cite{9,10,11}) have 
been developed. Recently in anyon optics, Laughlin states are made out of 
photon pairs (in a synthetic magnetic field for light induced from 
twisted optical cavity~\cite{19}, strong photonic interactions via 
Rydberg atoms). These are modeled exploiting  anyonic Hubbard Hamiltonian in ultracold-atom 1D 
lattices~\cite{pra}. Anyon imaging with STM has also been achieved~\cite{pra}. 
Theoretical~\cite{16,17} and experimental~\cite{18} studies for 
simulating anyonic NOON states with photons in waveguide lattices have appeared.\\
{\it {Our experimental proposal}}: Following the work in~\cite{18}, a two-photon 
NOON state with arbitrary anyonic symmetry is first prepared in a detuned directional
coupler, and subsequently evolved in a Bloch oscillator emulated by a curved array. 
Then we use {\em planar analogue} of a spiral phase plate where phase 
shift proportional to the path length of the waves passing through the plate will 
occur. Upon superposition this will generate the anyon beam. This is schematically 
depicted in Fig.~\ref{fig7}. The left panel of Fig.~\ref{fig7} shows a convex lens 
in three-dimensions $x,y,z$ with parallel rays shown only along $x$-$y$ plane that 
converge on $x$-axis. We consider an extremely thin slice of the lens in the $x$-$y$ 
plane which can be thought as the shaded area in the left panel. The same slice is 
drawn separately in the right panel that shows the planar superposition in our work.
  
\noindent
{(vi) \bf{Summary and future prospects:}} We have suggested the  construction of 
relativistic anyon beam, a symmetrical superposition of Jackiw-Nair single anyon 
solutions. Explicit forms of wave packets, their significant features and numerically 
plotted profiles of anyon beam current are given. A laboratory model of anyon 
beam construction is provided. 

We thank Prof. V. Parameswaran Nair for actively helping us in this project.

\vskip 0.6cm
\noindent
\rule[1cm]{\linewidth}{1pt}
\vskip -1cm
\noindent

\begin{center}
{\Large\bf{Supplemental material for ``Relativistic Anyon Beam: Construction 
and Properties"}}

\end{center}

\vskip 0.5cm
\noindent
{\bf{Supplemental material A: Explicit form of the free anyon wavefunction}}

\vskip 0.25cm
\noindent
In order to compute the free anyon wavefunction from eq.(7) of main text we require 
explicit form of $B_{n0}(p)$ sector of the infinite component matrix $B_{nn'}(p)$. 
The best strategy is to use coherent states, given in Ref.~\cite{jnsup}, replacing $z$ 
by a complex variable $\omega$. The coherent states are obtained by quantization of 
the canonical one-form
\begin{equation}
d\mathcal{A} = - i \lambda \left(\frac{\omega d \bar{\omega} - \bar{\omega} 
	d\omega}{1-\omega \bar{\omega}}\right) .
\label{ccan}
\end{equation}
These ideas have been discussed by Nair in \cite{vpnsup}. This one-form defines a symplectic structure on SU(1, 1)/U(1), where
\begin{equation}
g = 
\begin{pmatrix}
1 & \omega \\
\bar{\omega} & 1
\end{pmatrix}
\frac{1}{\sqrt{1-\bar{\omega}\omega}} \times 
\begin{pmatrix}
e^{i\phi/2} & 0 \\
0 & e^{-i\phi/2}
\end{pmatrix}
\end{equation}
The $\phi$-part of $g$ being irrelevant in $dA$, is not shown or equivalently we set $\phi=0$. (Construction of normalized coherent states of $SU(1,1)$ and its applications are discussed in \cite{persup}.) The quantization
of the canonical structure in (\ref{ccan}) gives the coherent states given in 
\cite{jnsup} (in terms of $\omega, \bar{\omega}$ rather than $z,z^*$). 
One can view these coherent states as being given by the matrix element  $g_{n0}\sim <\lambda, n\mid g \mid \lambda, 0>$. Here
$g_{nm}$ is the group element in the appropriate representation of $g$ obeying the normalization condition
$\sum_ng^*_{n0}g_{n0}=1$. Let us now develop the Fock space $\mid \lambda, n>$ in order to compute $g_{n0}$ explicitly. 

 Our first task is to write $g$ 
\begin{equation} 
g= \begin{pmatrix}
1&\omega \\
\bar{\omega}&1
\end{pmatrix} \frac{1}{\sqrt{1-\bar{\omega}\omega}}
\end{equation}
 in a factorized form. The $SU(1,1)\sim SO(2,1)$ group generators and their commutation rules are
given by 
\begin{equation}
[K_0,K_{\pm}]=\pm K_{\pm}, ~~~  [K_+,K_-]=-2K_0
\end{equation}
where $K_{\pm}=K_1 \mp iK_2$. A matrix realization in terms of  2$\times$2  Pauli matrices is
\begin{equation}
K_0=\frac{1}{2} \sigma_3, ~~~ K_1 = \frac{i}{2}\sigma_1, ~~~ K_2=-\frac{i}{2}\sigma_2 .
\end{equation}
Now it is straightforward to check that in the $2\times2$ matrix representation the following relation holds:
\begin{eqnarray}
e^{-i\omega K_+} e^{\log(1- \bar{\omega}\omega)K_0} e^{-i\bar{\omega}K_-} && = \exp \left[ \omega \begin{pmatrix} 
0&1 \\
0&0
\end{pmatrix} \right] e^{\sigma_3 \log(1-\bar{\omega} \omega)/2}  \exp \left[ \bar{\omega} \begin{pmatrix} 
0&0 \\
1&0
\end{pmatrix} \right] \nonumber \\
&& = \begin{pmatrix}
1&\omega \\0&1
\end{pmatrix} \begin{pmatrix}
\sqrt{1-\bar{\omega}\omega}&0 \\
0&\frac{1}{\sqrt{1-\bar{\omega}\omega}}
\end{pmatrix}\begin{pmatrix}
1&0 \\
\bar{\omega}&1
\end{pmatrix} \nonumber \\
&& =g .
\end{eqnarray}
Thus, we have managed to express $g$ as a group element in a conventional form as  the exponential of the generators with some parameters $\omega, \bar\omega $,
\begin{equation}
g= e^{-i\omega K_+} e^{\log(1- \bar{\omega}\omega)K_0} e^{-i\bar{\omega}K_-}.
\end{equation}
This is precisely  the infinite-dimensional bounded below representation of $g$~\cite{jnsup}
where the generators are to be taken in the representation of interest. 

Returning to the Fock space construction, explicit form of the  bounded below representation~\cite{jnsup} is given by 
\begin{eqnarray}
K^0\mid \lambda,n> = (\lambda + n) \mid\lambda,n>; \nonumber \\
K^+\mid\lambda , n> = \sqrt{(2\lambda+n)(n+1)}\mid\lambda,n+1>; \nonumber \\
K^-\mid \lambda,n>= \sqrt{(2\lambda+n-1)n}\mid\lambda,n-1> .
\label{km}
\end{eqnarray}
The lowest state of the representation is defined by 
\begin{equation}
K_-\mid \lambda,0 \rangle =0, ~~~ K_0 \mid \lambda, 0 \rangle = \lambda \mid \lambda,0 \rangle
\end{equation} 
with higher states are obtained by the action of powers of $K_+$ on $\mid 0 \rangle$ in an obvious way,  $K_0K_+^m \mid 0 \rangle = (\lambda + m ) K_+^m\mid 0 \rangle $. To recover the normalization we  define $f(n)$ (see e.g.~\cite{wysup}), 
\begin{eqnarray}
f(n) && = \langle \lambda,0 \mid K_-^n K_+^n \mid \lambda,0 \rangle \nonumber \\
&&  = \langle \lambda,0 \mid K_-^{n-1} [K_-,K_+^n ] \mid \lambda,0 \rangle\nonumber \\
&& = \langle \lambda,0\mid K_-^{n-1} (2K_0 K_+^{n-1} + K_+2K_0K_+^{n-2}+\cdot \cdot \cdot K_+^{n-1}2K_0) \mid \lambda,0 \rangle \nonumber \\
&& = 2 \sum_{k=1}^n (\lambda + n - k) f(n-1) \nonumber \\ 
&& = n(2\lambda + n -1) f(n-1) .
\end{eqnarray}
 Iterating and using $\langle 0 \mid 0 \rangle = 1 $, we find $f(n)$ in a closed form:
\begin{equation}
f(n) = n! (2\lambda + n - 1)(2\lambda + n - 2) \cdot \cdot \cdot (2\lambda) = \frac{\Gamma(n+1)\Gamma(2\lambda + n )}{\Gamma (2\lambda)}
\end{equation} 
where $\Gamma(u)$ is Eulerian gamma function for the argument u. The  normalized states are thus given by 
\begin{equation}
\mid \lambda, n \rangle = \frac{1}{\sqrt{f(n)}} K_+^n \mid \lambda,0 \rangle
\end{equation} 
Finally the cherished form of  matrix element $g_{n0}$ in this representation is derived, 
\begin{eqnarray}
g_{n0} && = \langle \lambda,n \mid e^{-i\omega K_+} e^{\log(1- \bar{\omega}\omega)K_0} e^{-i\bar{\omega}K_-}  \mid \lambda,0 \rangle \nonumber \\ 
&& = \langle \lambda,n \mid e^{-i\omega K_+} e^{\log(1- \bar{\omega}\omega)\lambda}  \mid \lambda,0 \rangle \nonumber \\ 
&& \langle \lambda,n \mid (-i \omega)^n \frac{K_+^n}{n!} \mid \lambda,0 \rangle (1- \bar{\omega} \omega)^{\lambda} \nonumber \\ 
&& = (-i\omega)^n (1-\bar{\omega}\omega)^{\lambda} \sqrt{\frac{\Gamma(2\lambda+n)}{\Gamma(2\lambda)n!}}
\end{eqnarray}
where we used $K_+^n \mid 0 \rangle = \sqrt{f(n)} \mid n \rangle $ from (12). A redefinition of variable ($-i\omega \rightarrow \omega$) 
yields, 
\begin{equation}
g_{n0} = \sqrt{\frac{\Gamma(2\lambda+n)}{n! \Gamma(2\lambda)}} (1-\bar{\omega}\omega)^{\lambda} \omega^n .
\label{eq20}
\end{equation} 
It is reassuring to check that the normalization condition mentioned earlier holds:
\begin{eqnarray}
(g^{\dagger}g)_{00}=
\sum_n g^*_{n0}g_{n0} && = \sum_n \frac{\Gamma(2\lambda +n)}{n! \Gamma(2\lambda)} (\bar{\omega}\omega)^n (1-\bar{\omega}\omega)^{2\lambda} \nonumber \\
&& = \frac{(1-\bar{\omega}\omega)^{2\lambda}}{\Gamma(2\lambda)} \sum_n \int_0^\infty dt e^{-t} \frac{t^{2\lambda+n-1}(\bar{\omega}\omega)^n}{n!} \nonumber \\
&& = \frac{(1-\bar{\omega}\omega)^{2\lambda}}{\Gamma(2\lambda)} \int_0^\infty dt e^{-t} e^{t\bar{\omega}\omega} t^{2\lambda-1} \nonumber \\
&& = \frac{(1-\bar{\omega}\omega)^{2\lambda}}{\Gamma(2\lambda)} \int_0^\infty dt e^{-t(1-\bar{\omega}\omega)}  t^{2\lambda-1} \nonumber \\
&& = 1 .
\end{eqnarray}
The $SU(1,1)\rightarrow SO(2,1)$ map: The manifestly  covariant field theoretic construction should reveal the single particle Poincare group representations of $SO(2,1)$~\cite{binsup}. In the present scheme $g$, being an element of SU(1,1) satisfies  $\sigma_3g^{\dagger}\sigma_3=g^{-1} $, or  equivalently $g^{\dagger}\sigma_3 g = \sigma_3$.  The adjoint representation of this is defined by 
\begin{equation}
R_{ij} = \frac{1}{2} Tr(g^{-1}K_{ig}K_j)
\end{equation}
with  $R_{ij}$  being real since it obeys $ K_i^{\dagger} \sigma_3 = \sigma_3 K_i $. Thus clearly $R_{ij}$ is   a real SO(2,1) matrix. This is the SU(1,1) $\rightarrow $ SO(2,1) map.

To construct $B_{n0}(p)$, we start with Eq.(3.6) of~\cite{jnsup},
\begin{equation}
B(p)=\frac{1}{\sqrt{2m}}[\sqrt{p.\eta+m}+i\frac{1}{\sqrt{p.\eta+m}}\epsilon^{abc}p_a\eta_b\gamma_c]
\label{01}
\end{equation}
with the convention 
\begin{equation}
\gamma^a=\{-\sigma^3,-i\sigma^2,i\sigma^1\};~\gamma_a=\{-\sigma^3,i\sigma^2,-i\sigma^1\};~\eta^a=\eta_a=\{1,0,0\},
\label{02}
\end{equation}
which gives $B_{n0}$ in the $2 \times 2$ matrix representation.

We use this $B(p)$ in place of $g$ in 
\begin{equation}
B(p)\equiv g = \sqrt{\frac{E+m}{2m}}
\left(\begin{array}{cc} 1 & -\frac{p_-}{E+m}\\ -\frac{p_+}{E+m} & 1 \end{array}\right)
\end{equation}
where $p_{\pm}=p_x\pm ip_y$. The above is written as  
\begin{equation}
B=\sqrt{\frac{1}{1-\omega\bar{\omega}}} \left(\begin{array}{cc} 1 & \omega\\ \bar{\omega} & 1 \end{array}\right)
\label{03}
\end{equation}
where $\omega=\frac{p_-}{E+m},~\bar{\omega}=\frac{p_+}{E+m}$.

Substituting $\omega$ in (\ref{eq20}) 
\begin{equation}
g_{n0}=\sqrt{\frac{\Gamma(2\lambda +n)}{n!\Gamma(2\lambda)}} \left(1-\omega\bar{\omega}\right)^\lambda \omega^n=
\sqrt{\frac{\Gamma(2\lambda +n)}{n!\Gamma(2\lambda)}} \left(\frac{2m}{E+m}\right)^\lambda 
	\left(\frac{p\textcolor{blue}{e^{-i\phi}}}{E+m}\right)^n
\label{04}
\end{equation}
where $p_x=p \cos\phi$, $p_y=p \sin\phi$ and $p=\sqrt{p_x^2+p_y^2}$.

The final expression is
\begin{equation}
g_{n0}=\sqrt{\frac{\Gamma(2\lambda +n)}{n!\Gamma(2\lambda)}}(2m)^\lambda (E+m)^{-(n+\lambda )}(pe^{-i\phi )})^n= B_{n0}.
\label{05}
\end{equation}

\vskip 0.3cm
\noindent
{\bf{Supplemental material B: Computational Details}}

\vskip 0.25cm
\noindent
In deriving the anyon current we have used the matrix representations
of $K^a$-matrices (see eq.(6) of main text)~\cite{jnsup},

\vskip 0.25cm
\noindent
Let us define, $K_{nn'}^a=\langle \lambda,n \mid K^a \mid \lambda,n' \rangle $ \\ Therefore,  $K_{nn'}^0=\langle \lambda,n \mid K^0 \mid \lambda,n' \rangle = (\lambda + n) \delta_{nn'}$, and \\
$K_{nn'}^+=\langle \lambda,n \mid K^+ \mid \lambda,n' \rangle = \langle \lambda,n \mid (\sqrt{(2\lambda+n')(n'+1)}) \mid \lambda , n'+1 \rangle = \sqrt{(2\lambda+n-1)n} \delta_{n,n'+1}$  \\ 
similarly,  $K_{nn'}^-=\langle \lambda,n \mid K^- \mid \lambda,n' \rangle = \sqrt{(2\lambda+n)(n+1)} \delta_{n,n'-1}  $ \\
Now, $K^{\pm} = K^x \mp iK^y$ \\
Therefore we finally have,
$$ K^x_{nn'}=\frac{1}{2} \left( K^+_{nn'} + K^-_{nn'} \right) = \frac{1}{2}\left(\sqrt{(2\lambda +n -1)n}\delta_{n,n'+1}
+ \sqrt{(2\lambda+n)(n+1)}\delta_{n,n'-1}\right) \nonumber $$ 

$$ K^y_{nn'}=\frac{i}{2} \left( K^+_{nn'} - K^-_{nn'} \right) =\frac{i}{2}\left(\sqrt{(2\lambda +n -1)n}\delta_{n,n'+1}-\sqrt{(2\lambda+n)(n+1)}\delta_{n,n'-1}\right)$$ 

\begin{eqnarray}
K^0_{nn'}=(\lambda + n) \delta_{nn'}
\end{eqnarray}
and the identity
\begin{equation}
\sum_{n=0}^{\infty}\frac{\Gamma(a+n)s^n}{n!} = (1-s)^{(-a)}\Gamma (a);~~ \mid s\mid < 1 .
\end{equation}

\noindent
{\bf{Supplemental material C: Expression for anyon beam current}}

\vskip 0.25 cm
\noindent
The spatial components of the anyon current built from the wavefunctions are:
\begin{eqnarray}
J^x(\rho, \theta) & = &  \left(\frac{2m}{E+m}\right)^{2\lambda} \frac{m}{E} \int\limits_{-\pi/2}^{\pi/2} 
d\alpha \int\limits_{-\pi/2}^{\pi/2} d\beta \left[e^{-2i \rho p \sin(\theta -(\alpha+ \beta)/2) \sin((\alpha-\beta)/2)} 
	\right.\nonumber \\ 
& & \left. + e^{-2i \rho p (\sin(\theta -(\alpha- \beta)/2) \sin((\alpha+\beta)/2)} + 
e^{-2i \rho p (\sin(\theta +(\alpha- \beta)/2) \sin(-(\alpha+\beta)/2)} \right. \nonumber \\ 
& & \left. + e^{-2i \rho p (\sin(\theta +(\alpha- \beta)/2) \sin((-\alpha+\beta)/2)}\right] \nonumber \\ 
& & \left[\frac{1}{2} \frac{p}{2m} \left\{ [1 - e^{-i (\alpha-\beta)} \sigma^2]^{-2 \lambda}  
\bigg( e^{-i\frac{\alpha}{2}}+e^{i \frac{\beta}{2}} + \frac{2p^2}{M} e^{-i(\alpha-\beta)} \cos(\alpha+\beta /2) 
\cos(\alpha-\beta/2)\bigg)\right\} \right. \nonumber \\ 
& & \left. - \frac{i \sigma \lambda}{4} \left\{ [1 - e^{-i (\alpha-\beta)} \sigma^2]^{-2 \lambda -1 }
\left(e^{-i\frac{\alpha }{2}} + e^{i\frac{\beta }{2}}\right) \left(-2i - 2i\frac{p^2}{M} + 
\frac{p^4}{M^2}e^{-i(\alpha-\beta)} \sin(\alpha-\beta)\right) \right\} \right].
\end{eqnarray}
and
\begin{eqnarray}
J^y(\rho, \theta) & = &  \left(\frac{2m}{E+m}\right)^{2\lambda} \frac{m}{E}  \int\limits_{-\pi/2}^{\pi/2} 
d\alpha \int\limits_{-\pi/2}^{\pi/2} d\beta \left[e^{-2i \rho p \sin(\theta -(\alpha+ \beta)/2) \sin((\alpha-\beta)/2)} 
	\right.\nonumber \\
& & \left. + e^{-2i \rho p (\sin(\theta -(\alpha- \beta)/2) \sin((\alpha+\beta)/2)} + 
e^{-2i \rho p (\sin(\theta +(\alpha- \beta)/2) \sin(-(\alpha+\beta)/2)}\right.\nonumber \\ 
& & \left. + e^{-2i \rho p (sin(\theta +(\alpha- \beta)/2)sin((-\alpha+\beta)/2)}\right] \nonumber \\ 
& & \left[\frac{1}{2} \frac{p}{2m} \left\{ [1 - e^{-i (\alpha-\beta)} \sigma^2]^{-2 \lambda}  
\left(i e^{-i\frac{\alpha}{2}}- i e^{i \frac{\beta}{2}} + \frac{2p^2}{M} e^{-i(\alpha-\beta)} 
\cos(\alpha+\beta /2) \cos(\alpha-\beta/2)\right)\right\} \right.\nonumber \\ 
& & \left.+ \frac{\sigma \lambda}{4} \left\{[1 - e^{-i (\alpha-\beta)} \sigma^2]^{-2 \lambda -1 }
\left(e^{-i\frac{\alpha }{2}} - e^{i\frac{\beta }{2}}\right) 
\left(-2i - 2i\frac{p^2}{M} + \frac{p^4}{M^2}e^{-i(\alpha-\beta)} \sin(\alpha-\beta)\right)\right\} \right].
\end{eqnarray}
where, $\sigma=p/(E+m)$.

Below we provide a few computational steps leading to $J^0$ given in (12).
$J^0$ is given by
\begin{equation}
J^0=\sum_{n=0}^{\infty} \left[F_n^{0\dagger}F^0_n+F_n^{x\dagger}F^x_n+F_n^{y\dagger}
F^y_n - i\left(F_n^{y\dagger}K^0_{nn'}F^x_{n'} - F_n^{x\dagger}K^0_{nn'}F^y_{n'}\right)\right].
\end{equation}

Consider the first term in the RHS where we have used expressions for the wave packet given in (12)
\begin{eqnarray}
\sum_{n=0}^{\infty} F_n^{0\dagger} F^0_n & = & \frac{1}{\Gamma(2\lambda)} \left(\frac{2m}{E+m}\right)^{2\lambda} 
\left(\frac{p}{2m} \right)^2 \left(\frac{m}{E}\right) \sum_{n=0}^{\infty}\frac{\Gamma(2\lambda+n)}{n!} 
\left(\frac{p}{E+m}\right)^{2n} \nonumber \\
& & \int\limits_{-\pi/2}^{\pi/2} \int\limits_{-\pi/2}^{\pi/2} d\alpha\, d\beta \,
	e^{-i n \alpha} e^{-i \alpha} \left[e^{-i( x \cos \alpha+y sin \alpha )} 
+ e^{-i( x \cos \alpha-y\sin \alpha )} \right] \nonumber \\
& & e^{i n \beta} e^{i \beta} \left[ e^{i( x \cos \beta+y \sin \beta )}
+ e^{i( x\cos \beta-y \sin \beta )} \right].
\end{eqnarray}

Now we substitute $x = \rho p \, \cos(\theta)$ and $y = \rho p \, \sin(\theta)$ and after 
using some well known trigonometric identity and the relation 
$\sum_{n=0}^{\infty} \frac{\Gamma(2\lambda + n )}{n!} x^n = (1-x)^{-2\lambda} \Gamma(2\lambda)$, 
we finally arrive at
\begin{eqnarray}
\sum_{n=0}^{\infty} F_n^{0\dagger} F^0_n & = &  \left(\frac{2m}{E+m}\right)^{2\lambda}\frac{m}{E} 
\int\limits_{-\pi/2}^{\pi/2} \int\limits_{-\pi/2}^{\pi/2} d\alpha\, d\beta \nonumber \\
& & \left[ e^{-2i \rho p \sin(\theta -(\alpha+ \beta)/2) \sin((\alpha-\beta)/2)} + 
     e^{-2i \rho p (\sin(\theta -(\alpha- \beta)/2) \sin((\alpha+\beta)/2)} \right. \nonumber \\ 
& & \left. +\, e^{-2i \rho p (\sin(\theta +(\alpha- \beta)/2) \sin(-(\alpha+\beta)/2)} 
     + e^{-2i \rho p (\sin(\theta +(\alpha- \beta)/2) \sin((-\alpha+\beta)/2)} \right] \nonumber \\ 
& & \left[1+e^{-i(\alpha-\beta)}\sigma^2\right]^{-2\lambda} \left(\frac{p}{2m}\right)^2 e^{-i(\alpha-\beta)} .
\end{eqnarray}

Now we write the expression for $J^0$ as below
\begin{equation}
J^0=\sum_{n=0}^{\infty} \left[F_n^{0\dagger}F^0_n+F_n^{x\dagger}F^x_n+F_n^{y\dagger}
F^y_n + i\lambda \left(F_n^{y\dagger}F^x_{n} - F_n^{x\dagger}F^y_{n}\right) - i (2\lambda + n)\left(F_n^{y\dagger}F^x_{n} - F_n^{x\dagger}F^y_{n}\right)\right].
\end{equation}
so that we can use the properties of Gamma function easily.
Similarly we can calculate $\sum_{n=0}^{\infty} F_n^{y\dagger} F^x_n$ as 
\begin{eqnarray}
\sum_{n=0}^{\infty} F_n^{y\dagger} F^x_n & = &  \left(\frac{2m}{E+m}\right)^{2\lambda}\frac{m}{E} 
\int\limits_{-\pi/2}^{\pi/2} \int\limits_{-\pi/2}^{\pi/2} d\alpha\, d\beta \nonumber \\
& & \left[ e^{-2i \rho p \sin(\theta -(\alpha+ \beta)/2) \sin((\alpha-\beta)/2)} + 
e^{-2i \rho p (\sin(\theta -(\alpha- \beta)/2) \sin((\alpha+\beta)/2)} \right. \nonumber \\ 
& & \left. +\, e^{-2i \rho p (\sin(\theta +(\alpha- \beta)/2) \sin(-(\alpha+\beta)/2)} 
+ e^{-2i \rho p (\sin(\theta +(\alpha- \beta)/2) \sin((-\alpha+\beta)/2)} \right] \nonumber \\ 
& & \left[1+e^{-i(\alpha-\beta)}\sigma^2\right]^{-2\lambda} \frac{1}{4} \left( -i -\frac{ip^2}{M} \cos(\beta) e^{i\beta} + \frac{p^2}{M} \sin(\alpha)e^{-i\alpha}+ \frac{p^4}{M^2}\sin(\alpha) \cos(\beta) e^{-i(\alpha-\beta)} \right).
\end{eqnarray}
and similarly the other terms of $J^0$ can be calculated. Putting all the terms  in Eq.(30) we find
\begin{eqnarray}
J^0(\rho, \theta) & = &  \left(\frac{2m}{E+m}\right)^{2\lambda} 
\frac{m}{E} \int\limits_{-\pi/2}^{\pi/2} d\alpha 
\int\limits_{-\pi/2}^{\pi/2} d\beta \left[e^{-2i \rho p \sin(\theta -(\alpha+ \beta)/2) \sin((\alpha-\beta)/2)} 
+ e^{-2i \rho p (\sin(\theta -(\alpha- \beta)/2) \sin((\alpha+\beta)/2)} \right.\nonumber \\ 
& & \left. + e^{-2i \rho p (\sin(\theta +(\alpha- \beta)/2) \sin(-(\alpha+\beta)/2)} +
e^{-2i \rho p (\sin(\theta +(\alpha- \beta)/2) \sin((-\alpha+\beta)/2)}\right] 
\left[\left[1 - e^{-i (\alpha-\beta)} \sigma^2\right]^{-2 \lambda} \right. \nonumber \\ 
& & \left. \left\{\left( \frac{p}{2m} \right)^2 e^{-i( \alpha-\beta)} 
+ \frac{1}{4} \left\{(1+\lambda)(2+ \frac{2p^2}{M}) + \frac{p^4}{M^2} e^{-i (\alpha-\beta)} (\cos(\alpha-\beta) 
+ i \lambda \sin(\alpha-\beta)) \right\} \right\} \right. \nonumber \\ 
& &\left. -\frac{\lambda}{2} \left\{ \left[1 - e^{-i (\alpha-\beta)} \sigma^2\right ]^{-2 \lambda -1 } 
\left(2 + 2\frac{p^2}{M} + i \frac{p^4}{M^2} \sin(\alpha-\beta)e^{-i(\alpha-\beta)}\right) \right\}\right].
\label{J0sup}
\end{eqnarray}

\vskip 0.3cm
\noindent
{\bf{Supplemental material D: A few more examples of  $J^0$ profiles  }}

\begin{figure}[ht]
	{\centering \resizebox*{10cm}{8cm}{\includegraphics{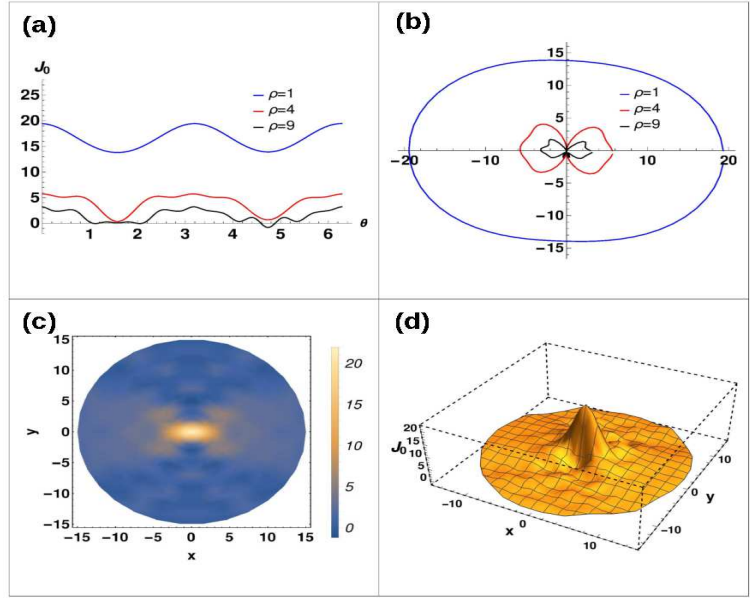}}\par}
	\caption{(Color online). Same as Fig. 2 of the main article with 
	$\lambda=0.2$.}
	\label{fig2sup}
\end{figure}

\begin{figure}[ht]
	{\centering \resizebox*{10cm}{8cm}{\includegraphics{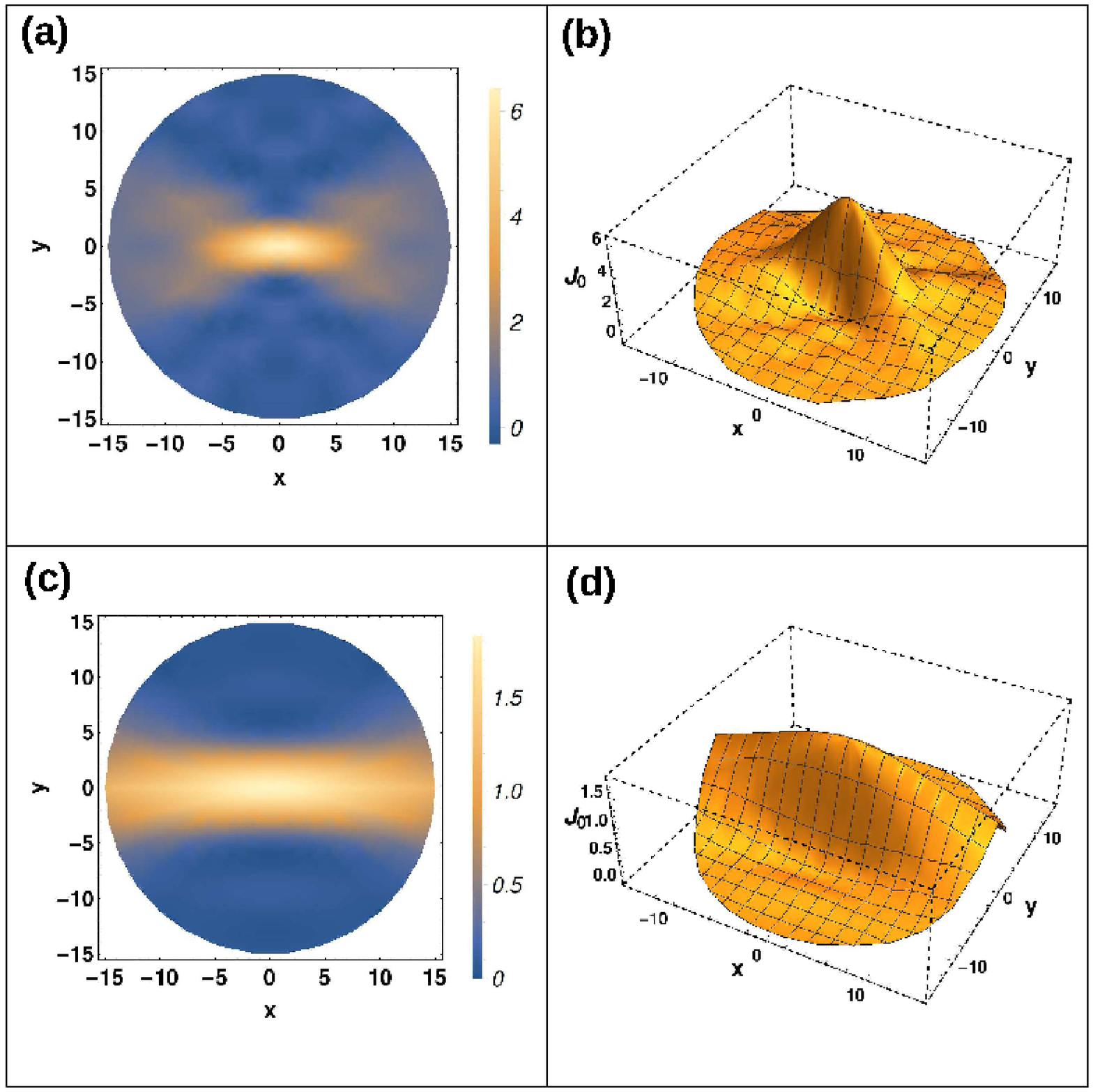}}\par}
	\caption{(Color online). Density plot and the $3$D plot of $J^0$ under two
		different integration ranges for superposition, considering $\lambda=0.6$. 
		In (a) and (b) we choose the integration range from $-\pi/3$ to $\pi/3$, 
		while in (c) and (d) the range is taken from $-\pi/6$ to $\pi/6$.}
	\label{fig4sup}
\end{figure}

\vskip 0.25cm
\noindent
In Fig.~\ref{fig2sup} we show the features of the anyon beam profile with 
numerical plots of $J^0$   
for $\lambda =0.2$ and superposition angle $\phi_0=\pm \pi/2$.  Again in Fig.~\ref{fig4sup} we have plotted $J^0$ 
for $\lambda =0.6$ for superposition angles having limiting values of 
$\phi_0=\pm\pi/3,~\pi/6 $. Description of the figures has already been given in the main text.

We thank Professor V. Parameswaran Nair for actively helping us in this
project.


\begin{thebibliography}{99}
	
\bibitem{dur} J. Durnin, J. Opt. Soc. Am. A \textbf{4}, 651 (1987); 
	
\bibitem{bl1} K. Y. Bliokh {\em et al.}, Phys. Rev. Lett. \textbf{99}, 190404 (2007); 
K. Y. Bliokh, M. R. Dennis, and F. Nori, Phys. Rev. Lett. \textbf{107}, 174802 (2011).
	
\bibitem{bl3} K. Y. Bliokh {\em et al.}, Phys. Rep. \textbf{690}, 1 (2017).

\bibitem{allen} L. Allen {\em et al.}, Phys. Rev. A \textbf{45}, 8185 (1992).

\bibitem{silenko} A. J. Silenko, P. Zhang, and L. Zou, Phys. Rev. Lett. \textbf{121}, 043202 (2019).
	
\bibitem{wil1} F. Wilczek, Phys. Rev. Lett. \textbf{49}, 957 (1982). 

\bibitem{lein} J. M. Leinaas and J. Myrheim , Nuovo Cimento \textbf{37B}, 1 (1977).
	
\bibitem{wil} {\em Fractional Statistics and Anyon Superconductivity}, World Scientific (1990), edited by F. Wilczek.	
	
\bibitem{graph} Y. Zhang {\em et al.}, Nature \textbf{438}, 201 (2005).
	
\bibitem{nonab gr} A. A. Zibrov {\em et al.}, Nature \textbf{549}, 360 (2017).
	
\bibitem{kit} A. Yu. Kitaev, Ann. Phys. (N. Y.) \textbf{303}, 2 (2003).

\bibitem{spin} H. Yao and S. A. Kivelson, Phys. Rev. Lett. \textbf{99}, 247203 (2007);	
M. Kapfer {\em et al.}, arXiv:1806.03117.
	
\bibitem{expt} L. Savary and L. Balents, Rep. Prog. Phys. \textbf{80}, 016502 (2017); 
S. Dutta and E. J. Mueller, Phys. Rev. A \textbf{97}, 033825 (2018).	
	 
\bibitem{jn} R. Jackiw and V. P. Nair, Phys. Rev. D \textbf{43}, 1933 (1991).

\bibitem{bin} B. Binegar, J. Math Phys. \textbf{23}, 1511 (1982).	

\bibitem{wy} B. G. Wybourne, {\em Classical Groups for Physicists}, Wiley, New York (1974).	

\bibitem{per} A. Perelomov, {\em Generalized coherent states and their applications}, Springer, Berlin (1986).
	
\bibitem{vpn} V. P. Nair, arXiv:1606.06407.
	
\bibitem{anyonobs} V. J. Goldman and B. Su, Science \textbf{267}, 1010 (1995);
V. J. Goldman, J. Liu, and A. Zaslavsky, Phys. Rev. B \textbf{71}, 153303 (2005).

\bibitem{obsan} F. E. Camino, W. Zhou, and V. J. Goldman, arXiv:cond-mat/0611443.
	
\bibitem{hos} Y. Hosotani, Int. J. Mod. Phys. B \textbf{7}, 2219 (1993);  
B. Abdullaev {\em et al.}, Phys. Rev. B \textbf{68}, 165105 (2003).
	
\bibitem{ch an} L. Yuan, M. Xiao, S. Xu, and S. Fan, Phys. Rev. A \textbf{96}, 043864 (2017).
	
\bibitem{wpacket} L. J. Krause et.al., Phys. Rev. Lett. \textbf{79}, 4978 (1997).

\bibitem{12} X. Du {\em et al.}, Nature \textbf{462}, 192 (2009).

\bibitem{13} K. I. Bolotin {\em et al.}, Nature \textbf{462}, 196 (2009).

\bibitem{14} E. M. Spanton {\em et al.} Science \textbf{360}, 62 (2018).

\bibitem{7} I. Bloch, J. Dalibard, W. Zwerger, Rev. Mod. Phys. \textbf{80}, 885 (2008).

\bibitem{8} N. R. Cooper, J. Dalibard, I. B. Spielman, Rev. Mod. Phys. \textbf{91}, 015005 (2019).

\bibitem{9} R. Umucallar, M.  Wouters, I. Carusotto, Phys. Rev. A \textbf{89}, 023803 (2014).

\bibitem{10} I. Carusotto, C.  Ciuti, Rev. Mod. Phys. \textbf{85}, 299 (2013).

\bibitem{11} T. Ozawa et al. Rev. Mod. Phys. \textbf{91}, 015006 (2019).

\bibitem{19} C. Noh and D. G. Angelakis, Rep. Prog. Phys. \textbf{80}, 016401 (2017).

\bibitem{pra} C. Yannouleas and U. Landman, Phys. Rev. A \textbf{100}, 013605 (2019)

\bibitem{16} Y. Bromberg, Y. Lahini, and Y. Silberberg, Phys. Rev. Lett. \textbf{105}, 263604 (2010).

\bibitem{17} S. Longhi and G. Della Valle,  Opt. Lett. \textbf{37}, 2160 (2012).

\bibitem{18} M. Lebugle {\em et al.}, Nat. Commun. \textbf{6}, 8273 (2015).
	
\end{thebibliography}

\begin{thebibliography}{99}

\bibitem{jnsup} R. Jackiw and V. P. Nair, Phys. Rev. D \textbf{43}, 1933 (1991).

\bibitem{vpnsup} V. P. Nair, {\em Elements of Geometric Quantization and Applications to Fields 
and Fluids}, arXiv:1606.06407.

\bibitem{persup} A. Perelomov, {\em Generalized Coherent States and Their Applications}, Springer, 
Berlin (1986); for a short introduction see M. Novaes, Revista Brasileira de Ensino de Fisica 
\textbf{26}, 351 (2004).

\bibitem{wysup} B. G. Wybourne, {\em Classical Groups for Physicists}, Wiley, New York (1974).

\bibitem{binsup} B. Binegar, J. Math Phys. \textbf{23}, 1511 (1982).	 
	
\end{thebibliography}
\end{document}